\documentclass[twocolumn, showpacs, preprintnumbers, amsmath, amssymb, superscriptaddress, pra]{revtex4}
\usepackage{graphicx}
\usepackage{dcolumn}
\usepackage{bm}
\usepackage{epsf}
\usepackage{amssymb}
\DeclareGraphicsExtensions{.eps}
\begin{document}

\author{David S. Simon}
\affiliation{Dept. of Electrical and Computer Engineering, Boston
University, 8 Saint Mary's St., Boston, MA 02215}
\affiliation{Dept. of Physics and Astronomy, Stonehill College, 320 Washington Street, Easton, MA 02357}

\author{Alexander V. Sergienko}
\affiliation{Dept. of Electrical and Computer Engineering, Boston
University, 8 Saint Mary's St., Boston, MA 02215}
\affiliation{Photonics Center, Boston
University, 8 Saint Mary's St., Boston, MA 02215}
\affiliation{Dept. of Physics, Boston University, 590 Commonwealth
Ave., Boston, MA 02215}

\begin{abstract} The concept of correlated two-photon spiral imaging is introduced. We begin by analyzing the joint orbital angular momentum (OAM) spectrum of correlated photon pairs. The mutual information carried by the photon pairs is evaluated, and it is shown that when an object is placed in one of the beam paths the value of the mutual information is strongly dependent on object shape and is closely related to the degree of rotational symmetry present. After analyzing the effect of the object on the OAM correlations, the method of correlated spiral imaging is described. We first present a version using parametric downconversion, in which entangled pairs of photons with opposite OAM values are produced, placing an object in the path of one beam. We then present a classical (correlated, but non-entangled) version. The relative problems and benefits of the classical versus entangled configurations are discussed. The prospect is raised of carrying out compressive imaging via two-photon OAM detection to reconstruct sparse objects with few measurements. \end{abstract}

\title{Two-Photon Spiral Imaging with Correlated Orbital Angular Momentum States}


\maketitle

\section{Introduction}

In digital spiral imaging (DSI) \cite{torres}, an object is illuminated by light with a known spatial distribution, or equivalently, of known orbital angular momentum (OAM) \cite{allen1,franke,yao} distribution, and the OAM spectrum after the object is measured. The shape of the outgoing spectrum allows determination of some properties of the object, or possibly identification of the object from a known set. However, as we will see below, this method is incapable of reconstructing the actual shape of the object; despite its name, it is inherently a {\it non-imaging} technique.

Here, we propose {\it correlated spiral imaging} (CSI), measuring correlations of OAM values within two-photon states or between two light beams. We will consider measurement of the correlations through both coincidence counting and through interference between the two beams. We will then show that (i) the CSI coincidence rate displays clear signatures of object spatial properties, (ii) the mutual information carried by the detected pair has a strong dependence on object shape and measures the object's rotational symmetry, and (iii) a version of the setup {\it does} allow efficient reconstruction of object shape, opening up the possibility of carrying out compressive imaging with high-dimensional OAM states.

The experiments proposed here differ significantly from that carried out in \cite{jack}. In the latter, after filtering for specific OAM values, the spatial locations of the outgoing photons are measured in one arm, as in traditional ghost imaging \cite{klyshko,belinskii,pittman,strekalov}. But in the present case, {\it no} information about the spatial location or momentum of the photon is recorded; {\it only} angular momentum values are detected.

In the following sections, we describe two categories of correlated spiral imaging experiments, one involving entangled photon pairs produced via downconversion, the other using classically correlated beams. The key point in all of the variations we describe is that there are {\it two} light beams (or two photons) with correlated OAM values. Our purposes here are two-fold: both scientific and applied. On the pure science side, the entangled version is of great interest, following in a direct line from work such as that of \cite{mair} and \cite{jack}, and offering a new window into both the downconversion process and the quantum correlations between the signal and idler OAM values. The latter correlations are certainly of scientific interest in their own right, apart from any applications. On the applied side, the classical version is likely to be more useful, as discussed more fully in section \ref{classicalsection}.

\section{Background}

\subsection{Laguerre-Gauss modes}
We decompose ingoing and outgoing beams in terms of optical Laguerre-Gauss (LG) modes.
The LG wavefunction with OAM $l\hbar$ and with $p$ radial nodes is \cite{allen2} \begin{eqnarray}& &u_{lp}(r,z,\phi )= {{C_p^{|l|}}\over {w(z)}}\left( {{\sqrt{2}r}\over {w(z)}}\right)^{|l|} e^{-r^2/w^2(r)}L_p^{|l|} \left( {{2r^2}\over {w^2(r)}}\right)\nonumber \\ & &\quad \times \; e^{ -ikr^2z/\left(2(z^2+z_R^2)\right)}e^{-i\phi l+i(2p+|l|+1)\arctan (z/z_R) },\label{lag}\end{eqnarray} with normalization $ C_p^{|l|} = \sqrt{ {{2p!}\over {\pi (p+|l|)!}}} $ and beam radius $w(z)= w_0\sqrt{1+{z\over {z_R}}} $ at $z$. $z_r={{\pi w_0^2}\over \lambda}$ is the Rayleigh range and the arctangent term is the Gouy phase.

\subsection{Digital spiral imaging}
\begin{figure}
\centering
\includegraphics[totalheight=1.8in]{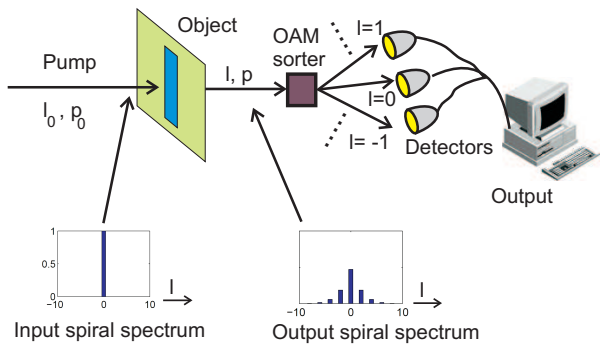}
\caption{\textit{Digital spiral imaging: the presence of an object in the light beam alters the distribution of angular momentum values in the outgoing light. }}\label{spiralfig}\end{figure}

DSI \cite{torres} is a form of angular momentum spectroscopy in which properties of an object are reconstructed based on how it alters the OAM spectrum of light used to illuminate it (fig. \ref{spiralfig}). The input and output light may be expanded in LG functions, with the object acting by transforming the coefficients of the ingoing expansion into those of the outgoing expansion. Information about the transmission profiles of both phase and amplitude objects may be retrieved \cite{torres,torres2}.

\begin{figure}
\centering
\includegraphics[totalheight=1.4in]{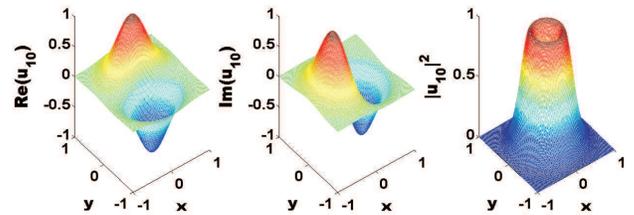}
\caption{\textit{The real and imaginary parts of the Laguerre-Gauss function are not rotationally-invariant, in contrast to its absolute square. This is illustrated for the case of $l=1$, $p=0$, but is true generally.}}
\label{laguerre10}\end{figure}

\begin{figure}
\centering
\includegraphics[totalheight=2.2in]{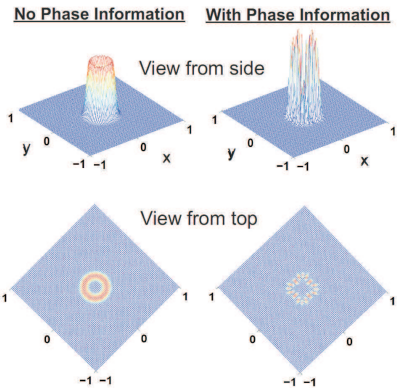}
\caption{\textit{Incoherent (left) and coherent (right) expansions in Laguerre-Gauss functions, an opaque square object. In the former case, all variation of the object with angle around the axis is lost. ($p_{max}=0$ and $l_{max}=15$ assumed.) }}
\label{square15}\end{figure}

The idea naturally arises of trying to use the measured OAM spectrum to reconstruct an image of the object. But, although a great deal of information may be obtained about the object in this manner, it is {\it not} sufficient to reconstruct a full image of the transmission or reflection profile. To see this, expand the output amplitude according to $\sum_{lp} A_{lp} u_{lp}$. Projecting out particular $l$ and $p$ values, the detector tells us the intensity of each component, allowing the $|A_{lp}|^2$ to be found, with no phase information retrieved.
We thus have an incoherent imaging setup, with total detected intensity of the form $\sum_{lp} |A_{lp}|^2 | u_{lp}|^2.$ But the quantities $| u_{lp}|^2$ are
rotationally symmetric for all values of $l$ and $p$ (see the right-most panel of fig. \ref{laguerre10}). Any image built from them is also symmetric; variation of the object about the axis is lost. In contrast, the real and imaginary parts are {\it not} rotationally invariant (left two panels of fig. \ref{laguerre10}), so a coherent sum of the form
$|\sum_{lp} A_{lp} u_{lp}|^2$ allows azimuthal structure to be reconstructed from the interference terms. For image reconstruction, we thus need to obtain a {\it coherent} superposition of amplitudes. This can be seen in fig. \ref{square15}: an opaque square is placed in the beam and the two expansions (coherent and incoherent) are computed, assuming that only the $p=0$ components are measured and keeping terms up to $|l_{max}|=15$. In the left panel, where no phase information is assumed, the reconstructed image is rotationally invariant and there is no way to distinguish what the actual shape of the object was. In contrast, the coherent expansion on the right side of the figure produces a recognizably square output. {\it The phase information is vital in reconstructing the actual image shape.}

\subsection{Entangled OAM beams}
Consider a pump beam of spatial profile $\Phi (\bm r)=u_{l_0p_0}(\bm r)$ encountering a $\chi^2$ nonlinear crystal, producing two outgoing beams via spontaneous parametric downconversion (SPDC). Assume a thin crystal located at the beam waist ($z=0$). The output is an entangled state \cite{mair}, with a superposition of terms of form $u_{l_1^\prime ,p_1^\prime }u_{l_2^\prime ,p_2^\prime }$, angular momentum conservation requiring $l_0=l_1^\prime +l_2^\prime $. We will take the pump to have $l_0=0$, so that the OAM values just after the crystal are equal and opposite: $l_1^\prime =-l_2^\prime \equiv l$. The $p_1^\prime, p_2^\prime$ values are unconstrained, although the amplitudes drop rapidly with increasing $p^\prime$ values (see eq. (\ref{Ccoefficient}) below).
The output of the crystal may be expanded as a superposition of signal and idler LG states:
\begin{equation}|\Psi\rangle =\sum_{l_1^\prime,l_2^\prime =-\infty }^\infty\sum_{p_1^\prime ,p_2^\prime =0}^\infty C^{l_1^\prime ,l_2^\prime }_{p_1^\prime p_2^\prime } |l_1^\prime ,p_1^\prime ;l_2^\prime ,p_2^\prime\rangle  \delta(l_0-l_1^\prime -l_2^\prime ) ,\end{equation} where the coupling coefficients are given by \begin{equation} C^{l_1^\prime ,l_2^\prime}_{p_1^\prime p_2^\prime}=\int d^2r \; \Phi (\bm r) \left[ u_{l_1^\prime p_1^\prime } (\bm r)u_{l_2^\prime p_2^\prime}(\bm r)\right]^\ast .\end{equation}
For the case of pump beam with $l_0=p_0=0$ this gives the coefficients \cite{torres2,ren}: \begin{eqnarray}C_{p_1,p_2}^{l,-l} &=&
\sum_{m=0}^{p_1}\sum_{n=0}^{p_2}\left( {2\over 3}\right)^{m+n+l}(-1)^{m+n} \label{Ccoefficient} \\
& & \times \; {{\sqrt{ p_1!p_2! (l+p_1)!(l+p_2)!}\; (l+m+n)!}\over {(p_1-m)!(p_2-n)!(l+m)!(l+n)!\; m!\; n!}} .\nonumber
\end{eqnarray}

\section{Joint OAM spectra}\label{OAMcorrelation}

\begin{figure}
\centering
\includegraphics[totalheight=1.8in]{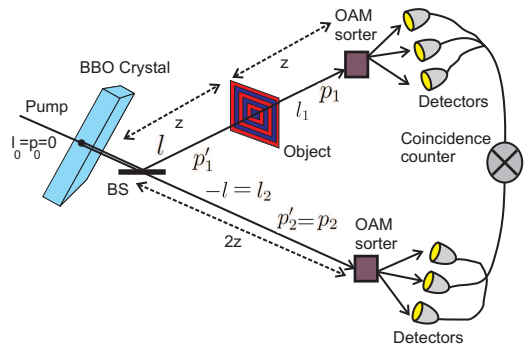}
\caption{\textit{Setup for analyzing object via orbital angular momentum of entangled photon pairs.}}
\label{entimage}\end{figure}

We now investigate the use of two beams, rather than one, in combination with spiral imaging. The full benefits of doing this will emerge in section \ref{imagingsection}. In the current section, we focus on examination of the OAM correlations. We begin with an entangled version, where the light source is parametric downconversion in a nonlinear crystal such as $\beta$-barium borate (BBO). Imagine an object in the signal beam (fig. \ref{entimage}). Since OAM conservation holds exactly only in the paraxial case, we assume the signal and idler are produced in {\it collinear} downconversion, then directed into separate branches by a beam splitter. (Throughout this paper we assume all beam splitters are 50-50.)
Assume perfect detectors for simplicity (imperfect detectors can be accounted for by the method in \cite{ren}).

Let $P(l_1,p_1;l_2,p_2)$ be the joint probability for detecting signal with quantum numbers $l_1,p_1$ and idler with values $l_2,p_2$.
The marginal probabilities at the two detectors (probabilities for detection of a single photon, rather than for coincidence detection) are \begin{eqnarray}P_s(l_1,p_1)&=&\sum_{l_2,p_2} P(l_1,p_1;l_2,p_2)\\ P_i(l_2,p_2)&=&\sum_{l_1,p_1} P(l_1,p_1;l_2,p_2).\end{eqnarray}
Then the mutual information for the pair is \begin{eqnarray}I(s,i)&=& \sum_{l_1,l_2=l_{min}}^{l_{max}} \sum_{p_1,p_2=0}^{p_{max}}P(l_1,p_1;l_2,p_2) \label{inf1}\\ & & \times \; \log_2 \left( {{P(l_1,p_1;l_2,p_2)}\over {P_s(l_1,p_1)P_i(l_2,p_2)}}\right)\nonumber \end{eqnarray} The most common experimental cases are when (i) the values of $p_1$ and $p_2$ are not measured (so all possible values of $p_1$ and $p_2$ must be summed, $p_{max}=\infty$ ), or (ii) only the $p_1=p_2=0$ modes are detected ($p_{max}=0$). Except when stated otherwise, we will use $l_{max}=-l_{min}=10$ and $p_{max}=0$.

If the transmission function for the object is $T({\bm x})$, the coincidence probabilities $P(l_1,p_1;l_2,p_2)=|A_{p_1p_2}^{l_1l_2}|^2$  have amplitudes \begin{eqnarray}A_{p_1p_2}^{l_1l_2}&=&C_0\sum_{ p_1^\prime}C_{p_1^\prime p_2}^{-l_2,l_2}a_{p_1^\prime p_1}^{-l_2, l_1}(z),\label{coherentamp}\\
a_{p_1^\prime p_1}^{l_1^\prime l_1}(z)&=&\int u_{l_1^\prime p_1^\prime }(\bm x_1,z) \left[ u_{l_1 p_1 }(\bm x_1,z)\right]^\ast T(\bm x_1)d^2x_1 \label{acoeff}\end{eqnarray} where $C_0$ is a normalization constant.
Here it is assumed that the total distance in each branch is $2z$ (see fig. \ref{entimage}).

\begin{figure}
\centering
\includegraphics[totalheight=1.2in]{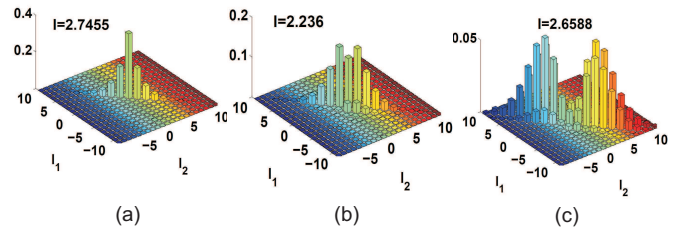}
\caption{\textit{An opaque strip of width $d$ placed in the signal path. The widths are (a) $d=.1 w_0$, (b) $d=.9 w_0$, (c) $d=2.5 w_0$. The outgoing joint angular momentum spectra are plotted. As the width increases, the peak in the spectrum broadens, then (at $d=w_0$) splits into two peaks.}}
\label{stripspectrumfig}\end{figure}

\begin{figure}
\centering
\includegraphics[totalheight=1.4in]{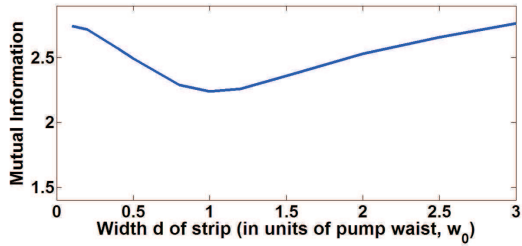}
\caption{\textit{Mutual information versus width of opaque strip. The horizontal axis is in units of $w_0$. The minimum information occurs at $d=w_0$. }}
\label{infowidthfig}\end{figure}

That the object's size and shape affect the coincidence rate is easy to see.
For example, fig. \ref{stripspectrumfig} shows the calculated spectrum when a single opaque strip of width $d$ is placed in the beam. Fig. \ref{infowidthfig} shows the corresponding mutual information, assuming that only the $p_1=p_2=0$ component is detected. In both figures, we see clear effects of changing an object parameter (the strip width).
The central peak of the spectrum (fig. \ref{stripspectrumfig}) broadens as $d$ increases from zero, reducing the correlation between $l_1$ and $l_2$; the mutual information between them thus declines, as seen in the $d/w_0<1$ portion of fig. \ref{infowidthfig}. But at $d/w_0\approx 1$, the central peak in $\left\{ l_1,l_1\right\}$ space bifurcates into two narrower peaks (right side of fig. \ref{stripspectrumfig}); the information thus goes back up as the peaks separate, as indeed is the case in the $d/w_0 >1$ region of fig. \ref{infowidthfig}. If we continue to wider $d$, the two peaks once again broaden and the mutual information decays gradually to zero. In addition, the total intensity getting past the opaque strip will continue to drop, so coincidence counts decay rapidly.

\section{Mutual information and symmetry}
\begin{figure}
\centering
\includegraphics[totalheight=1.0in]{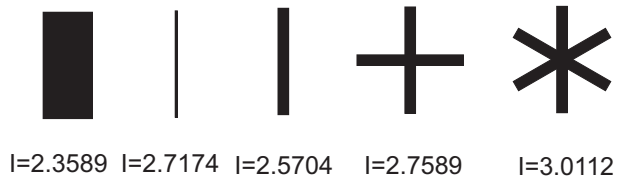}
\caption{\textit{The mutual information depends strongly on size and shape of the object. Here, the two objects on the left have widths $1.5w_0$ and $.2w_0$; all other widths are $.4w_0$.}}
\label{comparefig}\end{figure}

Fig. \ref{comparefig} shows the computed mutual information for several simple shapes. It can be seen that $I$ depends strongly on the size and shape of the object, so that for object identification from among a small set a comparison of the $I$ values rather than of the full probability distribution may suffice.

If the object has rotational symmetry about the pump axis, then its transmission function $T(r)$ depends only on radial distance  $r$, not on azimuthal angle $\phi$. The angular integral in eq. (\ref{acoeff}) is then
$\int_0^{2\pi} e^{-i\phi (l-l^\prime )}d\phi=2\pi \delta_{l,l^\prime}.$ So the joint probabilities reduce to $P(l_1,l_2)=f(l_1)\delta_{l_1,l_2}$ (assuming $p_1=p_2=0$). The marginal probabilities for each arm reduce to $P_1(l_1)= f(l_1)$ and $ P_2(l_2) =  f(l_2).$ The mutual information $ I(L_1,L_2)=S_1(L_1) $ where $S_1(L_1)=-\sum_{l_1}f(l_1)\ln f(l_1)$ is the Shannon information of the object arm OAM spectrum. Thus in the case of rotational symmetry, the second arm becomes irrelevant from an information standpoint. In this sense, the quantity $\mu (L_1,L_2) \equiv |I(L_1,L_2)-S_1(L_1)|$ is an order parameter, capable of detecting breaking of rotational symmetry.

More generally, suppose that the object has a rotational symmetry group of order $N$; i.e., it is invariant under $\phi\to \phi +{{2\pi}\over N}$. From eqs. (\ref{lag}) and (\ref{acoeff}) it follows that the coefficients must then satisfy $a_{p_1^\prime p_1}^{l_1^\prime l_1}=e^{{{2\pi i}\over N}(l_1^\prime -l_1)}a_{p_1^\prime p_1}^{l_1^\prime l_1}$, which implies $a_{p_1^\prime p_1}^{l_1^\prime l_1}=0$ except when ${{l_1^\prime -l_1}\over N}$ is integer. When $N$ goes up (enlarged symmetry group), the number of nonzero $a_{p_1^\prime p_1}^{l_1^\prime l_1}$ goes down; with the probability concentrated in a smaller number of configurations, correlations increase and mutual information goes up. This may be seen in the three right-most objects of fig. \ref{comparefig}, for example.

\subsection{Imaging.}\label{imagingsection}
The inability of DSI to produce images due to loss of phase information has been pointed out. Here we show that a variation on the entangled CSI setup can be used to find the expansion coefficients {\it including phase}.

First, we note from eq. (\ref{Ccoefficient}) that the factors $C^{l_1^\prime ,l_2^\prime}_{p_1^\prime p_2^\prime}$ are real and positive, so the phases of the amplitudes $A_{p_1p_2}^{l_1l_2}$ are entirely determined by the phases of the $a_{p_1^\prime p_1}^{-l_2, l_1}(z)$. Note further from eqs. (\ref{lag}) and (\ref{acoeff}) that the only $p$-dependence in the phase of $a_{p_1^\prime p_1}^{-l_2, l_1}(z)$ is in the factor $e^{-2ip\psi (z)}$, where $\psi (z)=\arctan {z\over {z_R}}$. If $z$ is much greater than $z_R$, then $\psi \approx {\pi\over 2}$ is roughly constant, so that $e^{-2ip\psi (z)}\approx (-1)p$. Since we assume the outgoing $p$ values are $p_1=p_2=0$, the relevant detection amplitude is \begin{equation}A_{00}^{l_1l_2} =\sum_{p_1^\prime}C_{p_1^\prime 0}^{-l_2,l_2}a_{p_1^\prime 0}^{-l_2,l_1} \approx C_{0 0}^{-l_2,l_2}a_{00}^{-l_2,l_1}+
C_{1 0}^{-l_2,l_2}a_{10}^{-l_2,l_1}, \end{equation} due to rapid decay of the $C_{p_1^\prime 0}^{-l_2,l_2}$ with increasing $|p_1^\prime -p_1|$. So if we break $a_{p_1^\prime 0}^{-l_2,l_1}$ into amplitude and phase, $a_{p_1^\prime 0}^{-l_2,l_1} =r_{p_1^\prime }^{-l_2,l_1} e^{i\phi_{p_1^\prime l_1l_2}}$, then the phase is independent of $p_1^\prime$, except for a relative minus sign between even and odd $p_1^\prime$ terms, so that \begin{equation}A_{00}^{l_1l_2}=\rho_{l_1,l_2}e^{i\phi_{l_1 l_2}},\end{equation} where $\phi_{l_1l_2}$ is the value of $\phi_{p_1^\prime l_1l_2}$ for even $p_1^\prime$, and $\rho_{l_1l_2}\equiv \left( C_{0 0}^{-l_2,l_2}-C_{1 0}^{-l_2,l_2}\right) r_0^{-l_2,l_1}$ is real and positive. Thus, the phase of $A_{00}^{l_1l_2}$ is the same as the phase of $a_{p_1^\prime 0}$ for even $p_1^\prime$ and differs from that of $a_{p_1^\prime 0}$ by a factor of $\pi$ for $p_1^\prime$ odd.
Finding the phases of the coincidence detection amplitudes $A_{00}^{l_1l_2}$ therefore suffices to determine the phases of all of the $a_{p_1^\prime 0}$ coefficients.

\begin{figure}
\centering
\includegraphics[totalheight=1.6in]{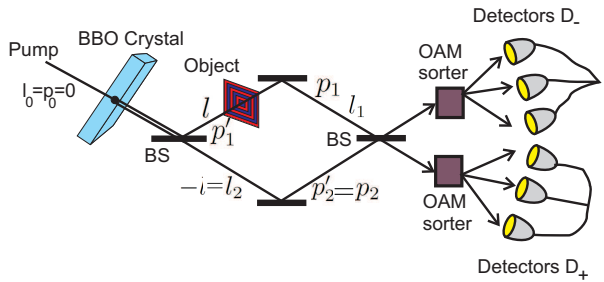}
\caption{\textit{A configuration allowing image reconstruction via phase-sensitive measurement of entangled OAM content.}}
\label{interferefig}\end{figure}

The measurement of these phases is accomplished by inserting a beam splitter to mix the signal and idler beams before detection, as in fig. \ref{interferefig}, erasing information about which photon followed which path. We then count singles rates in the two detection stages, rather than the coincidence rate.
If value $l$ is detected at a given detector it could have arrive by two different paths, so interference occurs between these two possibilities. The detection amplitudes in the two sets of detectors $D_+$ and $D_-$ involve factors $A_+\sim  \left( 1 + ia^{l_0-l_2,l_1}_{00}\right)$
and $A_-\sim  \left( i + a^{l_0-l_2,l_1}_{00}\right)$, with detection rates $R_\pm \sim  1 +|a^{l_0-l_2,l_1}_{00}|^2\pm 2i \; Im \; a^{l_0-l_2,l_1}_{00}.$  From these counting rates, both the amplitudes and the relative phases of all coefficients can be found, allowing full image reconstruction.

\section{Classical CSI}\label{classicalsection}

In recent years, it has been shown that ghost imaging and other "quantum" two-photon effects may be carried out using classically-correlated sources \cite{bennink1,bennink2,gatti2,valencia,scarcelli,ferri}.  It is apparent that the same is true in the case of correlated spiral imaging: classical OAM correlation, rather than entanglement, is sufficient.
The essential point in the present case is having two spatially separated beams such that if the OAM detected in one beam is known, then the OAM reaching the object can be predicted. So all that is needed is strong classical correlation or anti-correlation between the OAM in the two arms.

\begin{figure}
\centering
\includegraphics[totalheight=1.6in]{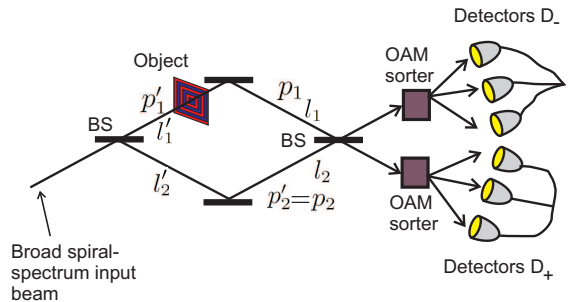}
\caption{\textit{A classical version of correlated spiral imaging. An input beam with a broad range of OAM values is split at a beam splitter, sending a portion through the reference branch, and the rest to the object.}}
\label{classicalfig}\end{figure}

The classical analog of apparatus of fig. \ref{interferefig} is shown in fig. \ref{classicalfig}. At the left, the system is illuminated with
light that has a broad range of OAM values (a broad spiral spectrum). The beam is split, with one copy passing through the object, and the other entering the reference branch. The two beams are mixed at the beam splitter, then the OAM content at the two detectors is measured. The coefficients $C_{p_1,p_2}^{l_1,l_2}$ will no longer be given by eq. \ref{Ccoefficient}, but instead will have values determined by the properties of the specific input beam being used. The mutual information between the classical beams may be defined just as in eq. \ref{inf1}.

The classical configuration of CSI has a number of practical advantages over the entangled version: alignment issues are greatly reduced, single photon detectors are not needed, and much higher brightness and counting rate may be obtained.
There is one problem that arises, however, which is not present in the entangled case: if a broad spiral spectrum is used for the illumination, then there is no intrinsic correlation between the OAM value $l_2$ in the reference branch and the value $l_1^\prime$ that occurs between the source and object. Without this correlation, the value of $l_1^\prime $ is unknown and so the change in $l$ produced by the object is also unknown. On the other hand, instead of a broad spiral spectrum, we may send in single OAM values, one at a time, building up the OAM correlation function one value of $l_1^\prime$ at a time.
But this slows the process of image reconstruction considerably: a range of OAM values needs to be scanned over, one after another, changing a spiral phase plate or some other type of OAM filter multiple times in each run. In a kind of quantum parallelism, the entangled version can send in a broad range of values simultaneously, and the entangled nature of the source will automatically ensure that the pairs detected are of opposite initial OAM if a short enough coincidence time window is used. In any case, whether the classical or entangled version is used, two correlated beams are necessary in order to reconstruct the relative phases of the various OAM amplitudes.

\section{Conclusion: Possible Extensions and Variations}

{\bf Compressive imaging.}
Recent years have shown an explosion of interest in compressive sensing \cite{candes1,candes2}, including compressive ghost imaging \cite{katz}. The basic idea is that most images are very
sparse when expanded in an appropriate basis, with the vast majority of expansion coefficients being very small. So if a sampling procedure is used that only measures the relatively small number of large expansion coefficients and neglects the rest, the image may be reconstructed
from a very small number of measurements, often much smaller than naively expected from the Shannon-Nyquist theorem.

The joint OAM spectra (such as those shown in fig. \ref{stripspectrumfig}) have been calculated for a variety of other opaque objects of various shapes, and in all of them it has been found that only a small number of the coefficients have significant amplitude; LG functions can therefore serve as a sparse basis for these shapes.  It is likely that this will also be true of at least some classes of more complex objects. The possibility thus opens of compressive imaging with OAM states by the CSI method. Only two additional ingredients are needed: (i) Instead of taking $l_0=0$, as we did in section \ref{OAMcorrelation}, we should illuminate the crystal with a broad range of $l_0$, providing a large number of randomly occurring input states; this provides the large number of randomly chosen sampling bases needed for compressive imaging. (ii) The basis used for sensing (LG basis, $|l,p\rangle$) and that used to reconstruct the image (position basis $|\bm r\rangle$) should have low mutual coherence \cite{candes1,candes2}; this means the maximum value of $|\langle \bm r|lp\rangle |=|u_{lp}(\bm r)| $ should be as small as possible, implying that the range of $l$ and $p$ values used should be centered at the largest possible mean value. A detailed discussion of this will be carried out elsewhere.

{\bf OAM Ghost imaging.} We have seen that the presence of an object yields clear signatures in both the CSI coincidence rate and the mutual information.
A variation to be explored elsewhere is "ghost" spiral imaging, where the $l_1$ values are not measured (analogous to having a bucket detector in ordinary ghost imaging). That signatures of the object will still appear can be seen by imagining summing over the $l_1$ rows in the histograms of fig. \ref{stripspectrumfig}.

This research was supported by  the DARPA InPho program through US Army Research Office award W911NF-10-1-0404.

\end{document}